\documentstyle[preprint,aps]{revtex}

\begin{document}
\title{Performing Quantum Measurement in Suitably Entangled States Originates the
Quantum Computation Speed Up}
\author{Giuseppe Castagnoli\thanks{%
Elsag Bailey and Universit\`{a} di Genova, 1654 Genova, Italy}, Dalida Monti%
\thanks{%
Elsag Bailey and Universit\`{a} di Genova, 1654 Genova, Italy}, and
Alexander Sergienko\thanks{%
Dept. of Electrical and Computer Engineering, Boston University, Boston, MA
02215, USA}}
\date{\today}
\maketitle

\begin{abstract}
We introduce a local concept of speed-up applicable to intermediate stages
of a quantum algorithm. We use it to analyse the complementary roles played
by quantum parallel computation and quantum measurement in yielding the
speed-up. A severe conflict between there being a speed-up and the many
worlds interpretation is highlighted.
\end{abstract}

\section{Introduction}

\noindent Why quantum computation can be more efficient than its classical
counterpart is a fundamental problem that has already attracted significant
attention (Ekert and Jozsa, 1997; Kitaev, 1997). The reason has naturally
been sought in the special features of quantum mechanics exploited in
quantum computation, like state superposition, entanglement and quantum
interference. For what concerns quantum measurement, until now it has been
thought to perform the following functions:

\begin{itemize}
\item  of course, accessing the outcome of the quantum computation process;

\item  selecting that outcome in a random way. In fact, in the case of some
algorithms, a certain number of {\em different} outcomes -- obtained through
repetition of the overall algorithm -- is needed to identify the problem
solution. Randomness assures that, by repeating the algorithm, we do not
always obtain the same outcome.
\end{itemize}

We will show that quantum measurement also plays a well defined role in
yielding the speed-up. This is due to the non-causal principle that the
measurement outcome is always a {\em single} eigenvalue of the measurement
basis. Given a suitable condition, this principle imposes a system of
algebraic equations representing the problem to be solved (or the hard part
thereof), whereas the measurement outcome, by satisfying that system, yields
the problem solution.

The condition is that the state before measurement is (suitably) entangled
with respect to the contents of two computer registers\footnote{%
In the current formulation of some quantum algorithms (e.g. the seminal 1985
Deutsch's algorithm and 1996 Grover's algorithm), this entanglement does not
appear. Likely, this has prevented understanding the role played by the
interplay between quantum parallel computation and quantum measurement in
yielding the speed-up. We will reformulate such algorithms by {\em physically%
} {\em representing} both the problem and the solution algorithm: in this
way the required entanglement appears.}. Measuring either content, through
the single outcome principle, imposes and solves the above mentioned
algebraic system.

The time required to measure the content of either register is linear in the
number of the register qubits, and it is not affected by registers
entanglement -- entanglement is interaction free. All the problem complexity
is represented in this entanglement and does not affect measurement time:
this is also essential to achieve the speed-up.

We shall examine all quantum algorithms found so far under the above
perspectives.

\section{Overview}

\noindent For the sake of clarity, we shall provide an overview of our
explanation of the speed-up, applied to a simplified version of Simon's
algorithm. All details are deferred to the subsequent Sections.

The problem is as follows. Given $B\sim \left\{ 0,1\right\} $ and $B^{n}\sim
\left\{ 0,1,...,N-1\right\} \sim {\bf Z}_{N}$ with $N=2^{n}$, we consider a
function $f\left( x\right) $ from $B^{n}$ to $B^{n}$. The argument $x$
ranges over $0,1,$ $...,$ $N-1$; $n$ is said to be the size of the problem. $%
f\left( x\right) $ has the following properties:

\begin{itemize}
\item  it is a 2-to-1 function, namely for any $x\in B^{n}$ there is one and
only one second argument $x^{^{\prime }}\in B^{n}$ such that $x\neq
x^{^{\prime }}$ and $f\left( x\right) =f\left( x^{^{\prime }}\right) $;

\item  such $x$ and $x^{^{\prime }}$ are evenly spaced by a constant value $%
r $, namely: $\ \left| x-x^{^{\prime }}\right| =r$;

\item  given a value $x$ of the argument, computing the corresponding value
of $f\left( x\right) $ requires a time polynomial in $n$ [i.e. poly$\left(
n\right) $]; whereas, given a value $f$ of the function, finding an $x$ such
that $f\left( x\right) =f$, requires a time exponential in $n$ [i.e. exp$%
\left( n\right) $] with any known classical algorithm; the function is
``hard to reverse''.
\end{itemize}

Besides knowing the above properties, the problem solver can access a
quantum computer that, given any input $x$, produces the output $f\left(
x\right) $ in poly$\left( n\right) $ time. The problem is to find $r$ in an
efficient way, which turns out to require poly$\left( n\right) $ time rather
than the exp$\left( n\right) $ time required by classical computation.

The computer operates on two registers $X$ and $F$, each of $n$ qubits; $X$
contains the argument $x$ and $F$ -- initially set at zero -- will contain
the result of computing $f\left( x\right) $. We denote by ${\cal H}%
_{XF}\equiv $ span$\left\{ \left| x\right\rangle _{X},\left| y\right\rangle
_{F}\right\} $, with $\left( x,y\right) $ running over $B^{n}\times B^{n}$,
the Hilbert space of the two registers. The initial state of these registers
is

\begin{equation}
\left| \varphi ,t_{0}\right\rangle _{XF}=\left| 0\right\rangle _{X}\left|
0\right\rangle _{F}
\end{equation}

They are, so to speak, blank. By using standard operations like the Hadamard
transform, and the quantum computer to perform function evaluation (see IV\
for details), we obtain in poly$\left( n\right) $ time the following
registers state (indexes are as in IV):

\begin{equation}
\left| \varphi ,t_{2}\right\rangle _{XF}=\frac{1}{\sqrt{N}}\sum_{x}\left|
x\right\rangle _{X}\left| f\left( x\right) \right\rangle _{F},
\end{equation}

\noindent with $x$ running over $0,1,$ $...,$ $N-1$. This is of course the
result of ``parallel quantum computation''.

Let us designate by $\left[ F\right] $ the number stored in register $F$ --
by $\left[ X\right] $ the number stored in $X$. We measure $\left[ F\right] $
in state (2) \footnote{%
This intermediate measurement can be skipped, but we will show (Section
IV.A) that either performing or skipping it is mathematically equivalent .
It has been introduced by Ekert and Jozsa (1998) to clarify the way Shor's
and Simon's algorithms operate; it also serves to clarify the speed-up.}.
Given the character of $f\left( x\right) $, the measurement outcome has the
form:

\begin{equation}
\left| \varphi ,t_{3}\right\rangle _{XF}=\frac{1}{\sqrt{2}}\left( \left| 
\overline{x}\right\rangle _{X}+\left| \overline{x}+r\right\rangle
_{X}\right) \left| \overline{f}\right\rangle _{F},
\end{equation}

\noindent where $\stackrel{\_}{f}$ is the value of the measured observable,
\noindent and $f\left( \overline{x}\right) =f\left( \overline{x}+r\right) =%
\overline{f}.$ The remaining part of the algorithm serves to extract $r$ out
of the above superposition, by using quantum interference within register $X$%
, measurement of $\left[ X\right] $ and repetition of the overall process
for a sufficient number of times (see Section IV.A).

However, we will see that, under a reasonable criterion, the speed-up has
already been achieved by preparing state (3), thus by performing the two
stages of quantum parallel computation and quantum measurement. How the two
stages interplay will be shown afterwards.

Since the speed-up is referred to an efficient classical computation that
yields the same result, the quantum character of state (3) constitutes a
difficulty. This is avoided by introducing a local definition of speed-up.
We should explicitly distinguish between a quantum state as a physical thing
and its symbolic description. By this we mean that, for example, the quantum
state described by (2) should be considered a preparation in a lab, while
expression (2), namely the ``string'' -- in the acception of formal
languages --

\noindent\ 
\[
\frac{1}{\sqrt{N}}\sum_{x}\left| x\right\rangle _{X}\left| f\left( x\right)
\right\rangle _{F}, 
\]

\noindent once all $x$ and $f\left( x\right) $ appearing in the sum are
substituted with the proper numerical values, is the symbolic description of
the preparation.

Let us consider the quantum computation cost (as a function of problem size)
of transforming a quantum state $A$ into a quantum state $B$. This is
benchmarked with the classical computation cost of transforming the symbolic
description of $A$ into the symbolic description of $B$. The speed-up is the
difference between the two. Cost is usually expressed as computation time
versus problem size, provided that all other computing resources are
comparable. This definition can be applied to intermediate stages of a
quantum algorithm, and coincides with the usual one if applied to the whole
algorithm.

First, we will use it to focus on stage (1) through (3) of Simon's
algorithm, in order to check that the speed-up, as defined above, originates
here. This will also serve to clarify the algebraic character of quantum
computation (Section III).

On the one hand, we should assess the time required to produce by classical
computation the symbolic description (3) starting from the symbolic
description (1) (i.e. from scratch). Naturally, $\overline{x},$ $\overline{x}%
+r,$ and $\overline{f}$ should be considered proper numerical values.
Finding them requires solving the following system of numerical algebraic
equations:

\begin{eqnarray}
f\left( x_{1}\right) &=&f\left( x_{2}\right) , \\
x_{1} &\neq &x_{2}.  \nonumber
\end{eqnarray}

\begin{center}
Fig. 1
\end{center}

It is useful to resort to the network representation of system (4) -- fig.
1. The gate $c\left( x_{1},x_{2}\right) $ imposes the condition that if $%
x_{1}\neq x_{2}$ then the output is 1, if $x_{1}=x_{2}$ then the output is
0, and vice-versa. To impose $x_{1}\neq x_{2}$, the gate output must be set
at 1. Note that this network is just a way of representing a system of
algebraic equations (useful to highlight a topological feature, as we will
see): time is not involved and gates are simply logical constraints. Each of
the two gates $f\left( x\right) $ imposes that if the input is $x$ then the
output is $f\left( x\right) $ or, conversely, if the output is $f$ then the
input is an $x$ such that $f\left( x\right) =f.$

This network is hard to satisfy by classical means. Because of the looped
network topology, finding a valuation of $x_{1},x_{2}$ and $f$ which
satisfies the network requires reversing $f\left( x\right) $ at least once,
which takes by assumption exp$\left( n\right) $ time.

On the other hand, the time required to produce the quantum state (3) with
Simon's algorithm, is the sum of the poly$\left( n\right) $ time required to
produce state (2) by means of parallel quantum computation, and the time
required to measure $\left[ F\right] $. This latter is independent of state
(2) entanglement and is simply linear in $n$, the number of qubits of
register $F$. The overall time is poly$\left( n\right) $. The speed-up -- as
defined above -- has already been achieved (in Section IV we will verify
from another standpoint that the remaining part of the algorithm cannot
possibly host any speed-up).

Let us see {\em how} the speed-up is originated by the interplay between
quantum parallel computation and quantum measurement. This will be shown at
a conceptual level herebelow, and formally in Section III.

We shall focus on stage (2) through (3). Analyzing the classical computation
cost of deriving the symbolic description (3) from (2), provides an
appropriate background for showing how quantum measurement operates.

Description (2) can be conveniently visualized as the {\em print-out} of the
sum of 2$^{n}$ tensor products (see eq. 2). Loosely speaking, two values of $%
x$ such that $f\left( x_{1}\right) =f\left( x_{2}\right) $, must be exp$%
\left( n\right) $ spaced. Otherwise such a pair of values could be found in
poly$\left( n\right) $ time by classically computing $f\left( x\right) $ on
a poly$\left( n\right) $ number of consecutive arguments.

The point is that the print-out would create a Library of Babel\footnote{%
From the story ``The Library of Babel'' by J.L. Borges.} effect. Even for a
small $n$, it would fill the entire galaxy with, say, $...$ $\left|
x_{1}\right\rangle _{X}\left| f\left( x_{1}\right) \right\rangle _{F}$ $...$
here, and $...$ $\left| x_{2}\right\rangle _{X}\left| f\left( x_{2}\right)
\right\rangle _{F}$ $...$ [such that $f\left( x_{1}\right) =f\left(
x_{2}\right) $] in Alpha Centauri. Finding such a pair of print-outs would
still require exp$\left( n\right) $ time.

By the way, this shows that a capability of directly accessing the so-called
``exponential wealth'' of parallel quantum computation, namely the above
print-out (the fact this is impossible is sometimes ``regretted'') would be
nullified by its ``exponential dilution''.

Let us see the way quantum measurement operates on the result of quantum
parallel computation. It {\em distills} the desired pair of arguments in a
time \ linear in $n$ (the number of qubits of register $F$). In fact, it
does more than randomly selecting one measurement outcome; by selecting a
single outcome{\em \ }-- a definite value of{\em \ }$f$ -- {\em it performs
a logical operation} {\em crucial for solving the problem }(i.e. selecting
the two values of $x$ associated with the value of that outcome). So to
speak, quantum measurement would perform as an ``exponentially efficient''
librarian. Its computational role, complementary to the production of the
parallel computation outputs, appears to be self-evident (given a proper
understanding of the context).

Noticeably, this way of operating of quantum measurement ``strongly''
violates the principle of causality (i.e. that all events have an antecedent
cause), in the following sense. It is not only the case that a measurement
outcome occurs at random (this would be the usual violation), it is also
computed -- and physically determined -- in a non-causal way.

This is due to the fact that a quantum evolution undergoing selective
measurement is affected by {\em both} the initial conditions and the final
condition that there is a single measurement outcome. This outcome is
therefore caused by both ends, which naturally violates the principle of
antecedent causes. This appears to be an interesting and completely new
physical notion, highlighted by a computational context. It will be further
expounded in Sections III and IV.

\section{Quantum algebraic computation}

\noindent We will show that quantum computation is an entirely new paradigm
where there is isomorphism between the {\em algebraic} {\em definition} of a
solution and its {\em physical determination}. This is unlike the usual
notion of algorithm, where the isomorphism is between the definition of a
sequential computational procedure and its dynamical implementation. As we
will see, in quantum computation the definition of a solution is
non-sequential and the corresponding physical determination is non-dynamical
in character.

Let us show that the state after measurement (3) (which we consider for
short to be the\ problem solution) solves a system of algebraic equations
imposed by measuring $\left[ F\right] $ in state (2). This is parallel to
solving, by classical computation, the system of algebraic equations (4).

It is useful to apply von Neumann's quantum measurement model. Besides
registers $X$ and $F$, we must consider the pointer of a measurement
apparatus that will give the result of measuring $\left[ F\right] $. Let $%
\left| \psi ,t_{2}\right\rangle _{XFP}=\left| \varphi ,t_{2}\right\rangle
_{XF}\left| 0\right\rangle _{P}$ be the state before measurement of the
whole system; $\left| 0\right\rangle _{P}$ denotes the pointer initial state.

von Neumann's model has two steps. The first, corresponding to the
measurement interaction, is a unitary evolution $U$ leading from the state
before measurement to a ``provisional description'' of the state after
measurement:

\[
\left| \psi ,t_{2}\right\rangle _{XFP}=\frac{1}{\sqrt{N}}\sum_{x}\left|
x\right\rangle _{X}\left| f\left( x\right) \right\rangle _{F}\left|
0\right\rangle _{P}\stackrel{U}{\longrightarrow } 
\]

\begin{equation}
\left| \psi ,t_{3}\right\rangle _{XFP}=\frac{1}{\sqrt{N}}%
\mathop{\textstyle\sum}%
_{i}\left( \left| x_{i}\right\rangle _{X}+\left| x_{i}+r\right\rangle
_{X}\right) \left| f_{i}\right\rangle _{F}\left| f_{i}\right\rangle _{P},
\end{equation}

\noindent with $f_{i}=f\left( x_{i}\right) =f\left( x_{i}+r\right) $ running
over all the values taken by $f\left( x\right) $. As stated before,
measurement time $t_{3}-t_{2}$ is linear in $n$. As well known, description
(5) yields the appropriate correlation between $\left[ F\right] $ and the
indication of the pointer, but it conflicts with the empirical evidence that
both (i.e. $f_{i}$) must have a definite measured value.

The second step of von Neumann's model can be seen as a {\em reinterpretation%
} of description (5): the tensor products appearing in the right hand of (5)
should be considered mutually exclusive measurement outcomes (still at the
same time $t_{3}$) with probability distribution the square modules of the
respective probability amplitudes\footnote{%
It is the same in decoherence theory, where the terms of a mixture become
mutually exclusive measurement outcomes.}. This yields a measurement outcome
of the form:

\[
\frac{1}{\sqrt{2}}\left( \left| \overline{x}\right\rangle _{X}+\left| 
\overline{x}+r\right\rangle _{X}\right) \left| \overline{f}\right\rangle
_{F}\left| \overline{f}\right\rangle _{P}, 
\]

\noindent where $\overline{f}$ is a definite value. In the following, we
will disregard the mathematically redundant factor $\left| \overline{f}%
\right\rangle _{P}$.

Note that this reinterpretation, as it is, does not involve the notion of
time and is ``transparent'' to dynamics, in the sense it does not affect
measurement time $t_{3}-t_{2}$. Interestingly, the selection of $\left| 
\overline{x}\right\rangle _{X}+\left| \overline{x}+r\right\rangle _{X}$,
essential to achieve the speed-up, stems from the reinterpretation, i.e. by
the non-causal principle that there must be a single measurement outcome.

This principle is usually applied in a procedural way: by performing a
sequence of typographic operations (in the acception of formal languages) on
the symbolic description of the state before measurement, we obtain the
symbolic description of a possible state after measurement.

This procedural approach conceals the algebraic character of the
reinterpretational step. In fact, by applying a system of algebraic
equations to an ``unknown'' vector $\left| \varphi \right\rangle _{XF}$ of
the Hilbert space of the two registers ${\cal H}_{XF}$, we obtain the same
result. The form of this vector is naturally $\left| \varphi \right\rangle
_{XF}=%
\mathop{\textstyle\sum}%
_{x,y}\alpha _{x,y}\left| x\right\rangle _{X}\left| y\right\rangle _{F}$,
where $\left( x,y\right) $ runs over $B^{n}\times B^{n}$, and $\alpha _{x,y}$
are complex variables independent of each other up to normalization: $%
\mathop{\textstyle\sum}%
_{x,y}\left| \alpha _{x,y}\right| ^{2}=1.$

Let $\left\{ \left| f\right\rangle _{F}\right\} $ be the set of the
eigenstates of register $F\ $and $\left| \overline{f}\right\rangle _{F}$ $%
\in $ $\left\{ \left| f\right\rangle _{F}\right\} $ be the eigenstate
selected by the quantum measurement of $\left[ F\right] $. The first
equation is

\begin{equation}
\left| \overline{f}\right\rangle _{F}\left\langle \overline{f}\right|
_{F}\left| \varphi \right\rangle _{XF}=\left| \varphi \right\rangle _{XF},
\end{equation}

\noindent \noindent $\left| \overline{f}\right\rangle _{F}\left\langle 
\overline{f}\right| _{F}$ is naturally the projector on the Hilbert subspace 
${\cal H}_{XF}^{\overline{f}}=$ span$\left\{ \left| x\right\rangle
_{X},\left| \overline{f}\right\rangle _{F}\right\} $ with $x$ running over $%
B^{n}$ and $\left| \overline{f}\right\rangle _{F}$ being fixed; a $\left|
\varphi \right\rangle _{XF}$ satisfying eq. (6) is a free linear combination
of all the tensor products of ${\cal H}_{XF}$ containing $\left| \overline{f}%
\right\rangle _{F}$;

\begin{equation}
\left| \left\langle \varphi \right. \right| _{XF}\left| \left. \varphi
,t_{2}\right\rangle _{XF}\right| \text{ must be maximum;}
\end{equation}

\noindent \noindent $\left| \varphi \right\rangle _{XF}$ satisfying
equations (6) and (7) becomes the projection of $\left| \varphi
,t_{2}\right\rangle _{XF}$ on ${\cal H}_{XF}^{\overline{f}}:$%
\[
\left| \varphi \right\rangle _{XF}=\sqrt{\frac{N}{2}}\left| \overline{f}%
\right\rangle _{F}\left\langle \overline{f}\right| _{F}\left| \varphi
,t_{2}\right\rangle _{XF}; 
\]
\noindent this means that $\left| \overline{f}\right\rangle _{F}$ has
``dragged'' all the tensor products of $\left| \varphi ,t_{2}\right\rangle
_{XF}$ containing it. The solution of the two equations is indeed the state
after measurement $\left| \varphi ,t_{3}\right\rangle _{XF}$ given by eq.
(3), formerly derived in the procedural way.

To sum up, solving the system of algebraic equations (6-7) is equivalent to
performing the reinterpretational step of von Neumann's model. This does not
affect the first step. In other words, performing the first step gives ``for
free'' (without incurring any further dynamical cost) the solution of (6-7)%
\footnote{%
In any way, the process of satisfying equations (6-7) must be comprised
within the measurement interaction, namely in the time interval $\left[
t_{2},t_{3}\right] $, which is linear in $n$.}. Under the current definition
of speed-up, solving equations (6-7) is equivalent to solving the system of
algebraic equations (4), namely the classically hard part of the problem.

This isomorphism between algebraic definition and physical determination of
a solution, blurs a long-standing distinction of mathematical logic between
the notions of ``implicit definition'' and ``computation''.

In problem solving, a problem implicitly defines its solution\footnote{%
If the problem admits no solution, we should consider the meta-problem
whether the problem admits a solution (which always admits a solution).}.
For example, let us consider factorization: given the known product $c$ of
two unknown prime numbers $x$ and $y$, the numerical algebraic equation $%
x\cdot y=c$ {\em implicitly} {\em defines} the values of $x$ and $y$ that
satisfy it. The system of algebraic equations (4) constitutes a similar
example.

In the classical framework, an implicit or algebraic definition of a
solution does not prescribe how to find it. In order to find the solution,
the implicit definition must be changed into an equivalent constructive
definition, namely into an algorithm (which is always possible with the
problems we are dealing with). An algorithm is an abstraction of the way
things can be constructed in reality -- inevitably in a model thereof -- and
prescribes a computation process that builds the solution.

The current notion of algorithm still reflects the way things can be
constructed in the traditional classical reality -- namely through a
sequential (dynamical/causal) process. Turing machine computation and the
Boolean network representation of computation are examples of sequential
(dynamical/causal) computation. In fact, an algorithm specifies a causal
propagation of logical implication from a completely defined input to a
completely defined output which contains the solution. It is thus meant to
be executable through a dynamical/causal process\footnote{%
Classical analog computation and classical nondeterministic computation are
not considered here to be fundamentally different, being still performed
through causal processes (classical randomness can be considered to be
pseudo-randomness). The important thing is that both differ from quantum
computation in their inability to host the non-causal effects we are dealing
with.}.

We can see that a feature of quantum computation essential to achieve the
speed-up, namely selective measurement, is extraneous to the causal notions
of both algorithm and dynamics. In particular, quantum computation {\em is
not} ``quantum Turing machine computation''. Out of habit, we will keep the
name of algorithm even in the case of a quantum computation yielding a
speed-up.

\section{Four types of quantum algorithms}

\subsection{\noindent Modified Simon's algorithm}

Without significant loss of generality, we will follow the simplified
version (Cleeve et al., 1997) of Simon's algorithm (1994). For the sake of
illustration, the following table gives a trivial example.

\begin{center}
\begin{tabular}{|c|c|}
\hline
$x$ & $f\left( x\right) $ \\ \hline
$0$ & $0$ \\ \hline
$1$ & $1$ \\ \hline
$2$ & $0$ \\ \hline
$3$ & $1$ \\ \hline
\end{tabular}

\bigskip

Table I

\medskip
\end{center}

\noindent The block diagram of the simplified algorithm is given in Fig. 2
-- we should disregard $/F$ for the time being.

\begin{center}
Fig. 2
\end{center}

\noindent The blocks in Fig. 2 represent the following transformations.

\begin{itemize}
\item  The $f\left( x\right) $ transform (a reversible Boolean gate in the
time-diagram of computation) leaves the content of register $X$ unaltered,
so that an input $x$ is repeated in the corresponding output, and computes $%
f\left( x\right) $ adding it to the former content of register $F$ (which
was set to zero). If the input state is not sharp but is a quantum
superposition, the same transformation applies to any tensor product
appearing in it.

\item  \medskip $H$ is the Hadamard transform defined as follows: $\left|
0\right\rangle _{i}\stackrel{H}{\longrightarrow }\frac{1}{\sqrt{2}}\left(
\left| 0\right\rangle _{i}+\left| 1\right\rangle _{i}\right) ,$ $\left|
1\right\rangle _{i}\stackrel{H}{\longrightarrow }\frac{1}{\sqrt{2}}\left(
\left| 0\right\rangle _{i}-\left| 1\right\rangle _{i}\right) $. In the case
of a register of n qubits, containing the number $\overline{x}$, it yields $%
\left| \overline{x}\right\rangle _{X}\stackrel{H}{\text{ }\longrightarrow 
\text{ }}\frac{1}{\sqrt{N}}\sum_{x}\left( -1\right) ^{\overline{x}\cdot
x}\left| x\right\rangle _{X},$ where $N=2^{n}$, $x$ ranges over $0,1,...$, $%
N-1$, and $\overline{x}\cdot x$ denotes the module 2 inner product of the
two numbers in binary notation (seen as row matrices).

\item  $M$ represents the action of measuring the numerical content of a
register.
\end{itemize}

We shall give some significant computational states (also for the example).
The preparation is $\left| \varphi ,t_{0}\right\rangle _{XF}=\left|
0\right\rangle _{X}\left| 0\right\rangle _{F}.$ After performing Hadamard on 
$X$ and function evaluation, we have:

\[
\left| \varphi ,t_{2}\right\rangle _{XF}=\frac{1}{\sqrt{N}}\sum_{x}\left|
x\right\rangle _{X}\left| f\left( x\right) \right\rangle _{F}=\frac{1}{2}%
\left( \left| 0\right\rangle _{X}\left| 0\right\rangle _{F}+\left|
1\right\rangle _{X}\left| 1\right\rangle _{F}+\left| 2\right\rangle
_{X}\left| 0\right\rangle _{F}+\left| 3\right\rangle _{X}\left|
1\right\rangle _{F}\right) ; 
\]
this is the state before the intermediate measurement of $\left[ F\right] $.

Measuring $\left[ F\right] $ yields, say, $\overline{f}=1$; the state after
measurement is consequently:

\[
\left| \varphi ,t_{3}\right\rangle _{XF}=\frac{1}{\sqrt{2}}\left( \left| 
\overline{x}\right\rangle _{X}+\left| \overline{x}+r\right\rangle
_{X}\right) \left| \overline{f}\right\rangle _{F}=\frac{1}{\sqrt{2}}\left(
\left| 1\right\rangle _{X}+\left| 3\right\rangle _{X}\right) \left|
1\right\rangle _{F}, 
\]

Ekert and Jozsa (1998) have shown that quantum entanglement between qubits
is essential for providing a speed-up, in terms of time or other computing
resources, in the class of quantum algorithms we are dealing with. After
measuring $f\left( x\right) $, the state of the two registers becomes
factorizable, and all entanglement is destroyed. The remaining actions,
performed on register $X$, use interference (which generates no
entanglement) to ``extract'' $r$ out of the superposition $\frac{1}{\sqrt{2}}%
\left( \left| \overline{x}\right\rangle _{a}+\left| \overline{x}%
+r\right\rangle _{a}\right) $. We must conclude from another standpoint that
the speed-up has been achieved by preparing $\left| \varphi
,t_{3}\right\rangle _{XF}$.

Performing again $H$ on register $X$ yields:

\[
\left| \varphi ,t_{4}\right\rangle _{XF}=\frac{1}{\sqrt{2N}}\sum_{z}\left(
-1\right) ^{\overline{x}\cdot z}\left[ 1+\left( -1\right) ^{r\cdot z}\right]
\left| z\right\rangle _{X}\left| \overline{f}\right\rangle _{F}. 
\]

Let us designate by $\overline{z}$ the result of measuring $\left[ X\right] $
in $\left| \varphi ,t_{4}\right\rangle _{XF}$; $r\cdot \overline{z}$ must be
0 -- see the form of $\left| \varphi ,t_{4}\right\rangle _{XF}$. This latter
property holds unaltered even if the intermediate measurement of $\left[ F%
\right] $ is omitted, as well known. By repeating the overall computation
process a sufficient number of times, poly($n$) on average, a number of
different constraints $r\cdot \overline{z}_{i}=0$ sufficient to identify $r$
is gathered.

How the speed-up is achieved in $\left[ t_{0},t_{3}\right] $ has been
anticipated in Sections II and III. Summarizing, measuring $\left[ F\right] $
in state $\left| \varphi ,t_{2}\right\rangle _{XF}$ imposes the system of
algebraic equations (6-7) [corresponding to (4)] and yields the
superposition of a pair of values of $x_{1}$ and $x_{2}$ which satisfy this
system ($r$ is ``easily'' extracted from this superposition). The overall
procedure requires poly(n) time, whereas solving equations (4) by classical
computation would require exp($n$) time.

Finally, let us show that performing or skipping the ``intermediate
measurement'' of $\left[ F\right] $ in $\left| \varphi ,t_{2}\right\rangle
_{XF}$, is {\em equivalent}. If we skip it (in fig. 2, the symbol $M\ $on $F$
should be erased), the state of registers $X$ and $F$ at time $t_{4}$
remains entangled. In the case of the example, this is 
\begin{equation}
\left| \varphi ,t_{4}\right\rangle _{XF}=\frac{1}{2}\left[ \left( \left|
00\right\rangle _{X}+\left| 01\right\rangle _{X}\right) \left|
0\right\rangle _{F}+\left( \left| 00\right\rangle _{X}-\left|
01\right\rangle _{X}\right) \left| 1\right\rangle _{F}\right]
\end{equation}

\noindent A direct measurement of register $X$ qubits is {\em equivalent} to
measuring first the content of $X$ with respect to the following basis:

\[
\left\{ \left| 00\right\rangle +\left| 01\right\rangle ,\left|
00\right\rangle -\left| 01\right\rangle ,\left| 10\right\rangle +\left|
11\right\rangle ,\left| 10\right\rangle -\left| 11\right\rangle \right\} . 
\]

\noindent In fact, one can see that introducing this previous measurement is
a {\em redundant }operation with respect to measuring the individual qubits.
Because of state (8) entanglement, this measurement is equivalent to
measuring $\left[ F\right] $ instead.

The result is, say:

\begin{equation}
\frac{1}{\sqrt{2}}\left( \left| 00\right\rangle _{X}-\left| 01\right\rangle
_{X}\right) \left| 1\right\rangle _{F}.
\end{equation}

\noindent

It is useful to resort to the notion of wave function collapse -- by the way
without having to adhere to it. We are just discussing a mathematical
equivalence, and the notion of collapse is a mathematically legitimate one.
According to von Neumann, collapse on state (9) can be backdated in time
provided that its result (9) undergoes in an inverted way the same
transformation undergone by the time-forward evolution (the usual one). It
can readily be seen that backdating this collapse at time $t_{2}$ is
equivalent to having performed the intermediate measurement of $\left[ F%
\right] $ in state (2).

Finally, it should be noted that $\left[ F\right] $ does not actually need
to be measured. This measurement is a virtual consequence of measuring only $%
\left[ X\right] $ at the end.

\subsection{Shor's algorithm}

\noindent The problem of factoring an integer $L$ -- the product of two
unknown primes -- is transformed into the problem of finding the period of
the function $f\left( x\right) =a^{x}%
\mathop{\rm mod}%
L$, where $a$ is an integer between $0$ and $L-1$, and is coprime with $L$
(Shor, 1994; Cleve et al., 1997). Figure 2 can also represent Shor's
algorithm, provided that $f\left( x\right) $ is as defined above and the
second Hadamard transform $H$ is substituted by the discrete Fourier
transform $F$. The state before measurement has the form $\left| \varphi
,t_{2}\right\rangle _{XF}=\frac{1}{\sqrt{L}}\sum_{x}\left| x\right\rangle
_{X}\left| f\left( x\right) \right\rangle _{F}$. Measuring or not measuring $%
f\left( x\right) $ in $\left| \varphi ,t_{2}\right\rangle _{XF}$ is still
equivalent. By measuring it, the above quantum state changes into the
superposition

\begin{equation}
\overline{k}\left( \left| \overline{x}\right\rangle _{X}+\left| \overline{x}%
+r\right\rangle _{X}+\left| \overline{x}+2r\right\rangle _{X}+...\right)
\left| \overline{f}\right\rangle _{F},
\end{equation}

\noindent where $f\left( \overline{x}\right) =f\left( \overline{x}+r\right)
=...=\overline{f}$, and $\overline{k}$ is a normalization factor.

The second part of the algorithm generates no entanglement and serves to
``extract'' $r$ in polynomial time, by using Fourier-transform interference
and auxiliary, off line, mathematical considerations. Under the current
assumptions, the quantum speed-up has been achieved by preparing state (10):
the discussion is completely similar to that of the previous algorithm.

\subsection{Deutsch's 1985 algorithm}

\noindent The seminal 1985 Deutsch's algorithm has been the first
demonstration of a quantum speed-up. In its usual form, this algorithm
yields a deterministic output, apparently ruling out the role of selective
measurement. A\ more careful examination will show that this is not the case.

Deutsch's algorithm, and more in general quantum oracle computing, is better
seen as a competition between two players. One -- the oracle -- produces the
problem, the other should produce the solution. Sticking to Greek tradition,
we shall call the former player Sphinx, the latter Oedipus.

The competition is formalized as follows. Both players have complete
knowledge of a set of functions $f_{k}:B^{n}\rightarrow B^{n}$ ($k$ labels
the elements of the set). They can also access a computer that, set in its $%
k $-th mode, computes $f_{k}\left( x\right) $ given any input $x\in B^{n}$.
The Sphinx chooses $k$ at random, sets the computer in its $k$-th mode and
passes it on to Oedipus. Oedipus knows nothing of the Sphinx' choice and
must efficiently find $k$ by testing the computer input-output behaviour. He
is naturally forbidden to inspect the computer mode. If the computer is
quantum, then we speak of ``quantum oracle computing''.

Deutsch's (1985) algorithm, as modified in (Cleve et al., 1997), is as
follows. Let \noindent $\left\{ f_{k}\right\} $ be the set of all possible
functions \noindent $f_{k}:B\rightarrow B$, namely:

\medskip

\begin{center}
\begin{tabular}{|c|c|cc|c|c|cc|c|c|cc|c|c|}
\cline{1-2}\cline{5-6}\cline{9-10}\cline{13-14}
$x$ & $f_{00}\left( x\right) $ & \qquad &  & $x$ & $f_{01}\left( x\right) $
& \qquad &  & $x$ & $f_{10}\left( x\right) $ & \qquad &  & $x$ & $%
f_{11}\left( x\right) $ \\ \cline{1-2}\cline{5-6}\cline{9-10}\cline{13-14}
$0$ & $0$ & \qquad &  & $0$ & $0$ & \qquad &  & $0$ & $1$ & \qquad &  & $0 $
& $1$ \\ \cline{1-2}\cline{5-6}\cline{9-10}\cline{13-14}
$1$ & $0$ & \qquad &  & $1$ & $1$ & \qquad &  & $1$ & $0$ & \qquad &  & $1 $
& $1$ \\ \cline{1-2}\cline{5-6}\cline{9-10}\cline{13-14}
\end{tabular}
\end{center}

\noindent \medskip

\noindent $\left\{ f_{k}\right\} $ is divided into a couple of subsets: the
balanced functions, characterized by an even number of zero and one function
values, thus labeled by $k=01,10$, and the unbalanced ones, labeled by $%
k=00,11$. Oedipus must find, with a minimum number of computer runs, whether
the computer (whose mode has been randomly set by the Sphinx) computes a
balanced or an unbalanced function. The algorithm is illustrated in Fig.
3(a). The computation of $f_{k}\left( x\right) $ is represented as a
reversible Boolean gate, like in the previous algorithms but for the fact
that the result of the computation is now module 2 added to the former
content of register $F$.

\begin{center}
Fig. 3(a),(b)
\end{center}

Let us consider the significant computational states. The preparation is the
same for all $k$: $\left| \varphi _{k},t_{0}\right\rangle _{av}=\frac{1}{%
\sqrt{2}}\left| 0\right\rangle _{a}\left( \left| 0\right\rangle _{v}-\left|
1\right\rangle _{v}\right) .$The state at time $t_{3}$ before measurement,
after performing function evaluation by running the computer only once (Fig.
3a), depends on the Sphinx' choice:

\vspace{0.7cm}

if $k=00$ then $\left| \varphi _{00},t_{3}\right\rangle _{XF}=\frac{1}{\sqrt{%
2}}\left| 0\right\rangle _{X}$ $\left( \left| 0\right\rangle _{F}-\left|
1\right\rangle _{F}\right) $

if $k=01$ then $\left| \varphi _{01},t_{3}\right\rangle _{XF}=\frac{1}{\sqrt{%
2}}\left| 1\right\rangle _{X}$ $\left( \left| 0\right\rangle _{F}-\left|
1\right\rangle _{F}\right) $

if $k=10$ then $\left| \varphi _{10},t_{3}\right\rangle _{XF}=-\frac{1}{%
\sqrt{2}}\left| 1\right\rangle _{X}$ $\left( \left| 0\right\rangle
_{F}-\left| 1\right\rangle _{F}\right) $

if $k=11$ then $\left| \varphi _{11},t_{3}\right\rangle _{XF}=-\frac{1}{%
\sqrt{2}}\left| 0\right\rangle _{X}$ $\left( \left| 0\right\rangle
_{F}-\left| 1\right\rangle _{F}\right) $

\vspace{0.7cm}

It can be seen that measuring $\left[ X\right] $ gives Oedipus' answer (1
for balanced, 0 for unbalanced).

This algorithm is more efficient than any classical algorithm, where two
runs of the computer would be required to establish the answer. However, the
result is apparently reached in a deterministic way, without quantum
measurement performing any selection.

This must be ascribed to the incomplete physical representation of the
situation. In Sections IV.A and IV.B, we had a problem that implicitly
defined its solution. The physical determination of the solution was
obtained by measuring the content of a computer register in an entangled
state representing the problem to be solved (or the hard part thereof). This
obviously requires that the problem is physically represented\footnote{%
In Sections IV.A and IV.B, all knowledge of the function and ignorance about 
$r$ were physically represented in a superposition of the form (2).},
whereas presently an essential part of it, the Sphinx choosing the oracle
mode, is not. For example, the logical if-then conditions above are external
to the physical representation.

In order to complete the physical representation, we introduce the extended
gate $F\left( k,x\right) $ which computes the function $F\left( k,x\right)
=f_{k}\left( x\right) $ for all $k$ and $x$. This gate has an ancillary
input register $K$ which contains $k$, namely the oracle mode [Figure 3(b)
gives the extended algorithm]. This input is identically repeated in a
corresponding output -- to keep gate reversibility. Of course, Oedipus is
forbidden to inspect register $K$. The preparation becomes 
\[
\left| \varphi ,t_{0}\right\rangle _{KXF}=\frac{1}{2\sqrt{2}}\left( \left|
00\right\rangle _{K}+e^{i\delta _{1}}\left| 01\right\rangle _{K}+e^{i\delta
_{2}}\left| 10\right\rangle _{K}+e^{i\delta _{3}}\left| 11\right\rangle
_{K}\right) \left| 0\right\rangle _{X}\left( \left| 0\right\rangle
_{F}-\left| 1\right\rangle _{F}\right) . 
\]

\noindent where $\delta _{1}$, $\delta _{2}$ and $\delta _{3}$ are
independent random phases. The superposition $\left| 00\right\rangle
_{K}+e^{i\delta _{1}}\left| 01\right\rangle _{K}+e^{i\delta _{2}}\left|
10\right\rangle _{K}+e^{i\delta _{3}}\left| 11\right\rangle _{K}$ represents
the fact that the Sphinx gives Oedipus the computer in a mode $k$ randomly
chosen among $k=00,01,10,11$. To Oedipus, this is indistinguishable from a
mixture -- the above superposition is in fact a mixture represented with the
method of random phases (Finkelstein, 1996).

Let us go directly to the state before measurement -- see Fig. 3(b):

\begin{equation}
\left| \varphi ,t_{3}\right\rangle _{KXF}=\frac{1}{2\sqrt{2}}\left[ \left(
\left| 00\right\rangle _{K}-e^{i\delta _{3}}\left| 11\right\rangle
_{K}\right) \left| 0\right\rangle _{X}+\left( e^{i\delta _{1}}\left|
01\right\rangle _{K}-e^{i\delta _{2}}\left| 10\right\rangle _{K}\right)
\left| 1\right\rangle _{X}\right] \left( \left| 0\right\rangle _{F}-\left|
1\right\rangle _{F}\right) .
\end{equation}

It can be seen that this entangled state represents the {\em mutual
definition} between the Sphinx' choice of the computer mode $k$ and Oedipus'
answer. The problem implicitly defining its solution appears here in the
form of the mutual definition of the moves of the two players. Reaching
state (11) still requires one oracle run.

The action of measuring $\left[ K\right] $ in state (11), equivalent to the
Sphinx' choice of the oracle mode, by imposing equations (6-7)\footnote{%
Of course, $\left| \varphi ,t_{2}\right\rangle _{XF}$ of Section III must be
changed into $\left| \varphi ,t_{3}\right\rangle _{KXF}$ and ${\cal H}_{XF}$
into ${\cal H}_{KXF}$.}, transforms mutual definition into correlation
between individual measurement outcomes (like in an EPR situation). In other
words, the Sphinx' choice of $k$ {\em simultaneously} determines the problem
solution, namely Oedipus' answer -- retrievable by measuring $\left[ X\right]
$. In the classical framework instead, the Sphinx' choice should necessarily
be propagated to Oedipus' answer by means of a causal process (requiring two
computer runs).

Achieving the speed-up still involves measurement of an entangled state,
namely selective measurement.

\subsection{An instance of Grover's algorithm}

\noindent The rules of the game are the same as before. This time we have
the set of the $2^{n}$ functions $f_{k}:B^{n}\rightarrow B$ such that $%
f_{k}\left( x\right) =\delta _{k,x}$, where $\delta $ is the Kronecker
symbol. We shall consider the simplest instance $n=2$. This yields four
functions $f_{k}\left( x\right) $, labeled $k=0,1,2,3$. Figure 4(a) gives
Grover's algorithm (1996) in the standard version provided in (Cleve et al.,
1997) for $n=2$. Without entering into detail, the complete physical
representation of the state before measurement becomes (Fig. 4b):

\begin{center}
Fig. 4(a),(b)
\end{center}

\[
\frac{1}{2\sqrt{2}}\left( \left| 0\right\rangle _{K}\left| 0\right\rangle
_{X}+e^{i\delta _{1}}\left| 1\right\rangle _{K}\left| 1\right\rangle
_{X}+e^{i\delta _{2}}\left| 2\right\rangle _{K}\left| 2\right\rangle
_{X}+e^{i\delta _{3}}\left| 3\right\rangle _{K}\left| 3\right\rangle
_{X}\right) \left( \left| 0\right\rangle _{F}-\left| 1\right\rangle
_{F}\right) , 
\]
where $\delta _{1}$, $\delta _{2}$ and $\delta _{3}$ are independent random
phases.

Again, we have the mutual definition of the Sphinx' choice and Oedipus'
answer. This is transformed into physical determination by measuring the
content of either register, as in the previous oracle problem.

\section{Conclusions}

\noindent This work should provide a better understanding of what quantum
computation is and is not at a fundamental level. It is not, as often
believed, the quantum transposition of sequential-causal computation (e.g.
of Turing machine computation). This is only the first stage of the quantum
algorithm. In the measurement stage, there is isomorphism between the
implicit definition of the problem solution and its (non-causal) physical
determination. The speed-up essentially relies on the non-causal principle
that the measurement oucome is a single eigenvalue of the measurement basis.

Detaching quantum computation from the notion of causal algorithm -- a
classical vestige -- should be a precondition for pursuing further
developments at a fundamental level.

For example, let us consider the possibility of exploiting particle
statistics symmetrizations to achieve a quantum speed-up. Such
symmetrizations can be seen as projections on symmetric (constrained)
Hilbert subspaces. There is no relation between a projection and reversible
Turing machine computation, namely a unitary evolution. If instead quantum
computation is (properly) seen as a projection on a constrained Hilbert
subspace, which amounts to solving a problem, then we have an analogy with
particle statistics symmetrizations to work with. Refs. (Castagnoli, 1998;
Castagnoli et al., 1998) provide still abstract attempts in this direction.

More generally, this work highlights the essential role played by
non-dynamical effects in quantum computation. Let us mention in passing that
a form of quantum computation which is of geometric rather than dynamic
origin has recently been provided (Jones et al., 2000). This concretely
shows that there are ways of getting out of the usual quantum computation
paradigm.

This work conflicts (Castagnoli et al., 2000) with the many worlds
interpretation. As well known, this interpretation aims to restore the
principle of causality by denying the objectivity of the (non-causal)
principle that there is always a single measurement outcome (denoted by $%
{\cal S}$ in the following). The fact we experience ${\cal S}$, would be
subjective in character -- ascribable to a limitation of our perception.

Thus, from the one hand, ${\cal S}$ would be subjective in character; from
the other, it yields the speed-up, an objective consequence in a most
obvious way. To avoid this contradiction, we are obliged to accept the
objectivity of ${\cal S}$. It can be argued that the ``strong'' form of
causality violation highlighted in this work (Section II) is too strong to
be denied by the many worlds interpretation.

\smallskip

Thanks are due to T. Beth, A. Ekert, D. Finkelstein and V. Vedral for
stimulating discussions and valuable comments.

\newpage

{\LARGE References}

Castagnoli, G. (1998), {\em Physica} {\em D} {\bf 120}, 48.

Castagnoli, G., and Monti, D. (1998). To be published in {\em Int. J. Theor.
Phys.};{\em \ }quant-ph/9811039.

Castagnoli, G., Ekert, A. (2000). Private discussion.

Cleve, R., Ekert, A., Macchiavello, C., and Mosca, M. (1997). Submitted to 
{\em Proc. Roy. Soc. Lond. A}; quant-ph/9708016.

Deutsch, D. (1985), {\em Proc. of the Royal Society of London A}, {\bf 400},
97.

Ekert, A., and Jozsa, R. (1998). {\em Proc. of Roy. Soc. Discussion Meeting}%
, to appear in {\em Phil. Trans. Roy. Soc. (London)}; quant-ph/9803072.

Finkelstein, D.R. (1996), ``Quantum Relativity'', {\em Springer-Verlag
Berlin Heidelberg}.

Grover, L. (1996), {\em Proc. of 28th Annual ACM\ Symposium on Theory of
Computing}.

Jones, J.A., Vedral, V., Ekert, A., and Castagnoli, G. (1999). {\em Nature}, 
{\bf 403}, 869; quant-ph/9910052.

Kitaev, A.Yu. (1997); quant-ph/9707021.

Shor, P. (1994). {\em Proc. of the 35}$^{th}${\em \ Annual Symposium on the
Foundation of Computer Science}, Los Alamitos, {\bf CA}, 124.

Simon, D.R. (1994). {\em Proc. of the 35}$^{th}${\em \ Annual Symposium on
the Foundation of Computer Science}, Santa Fe, IVM.

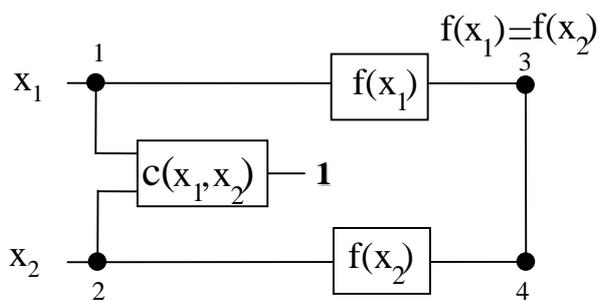
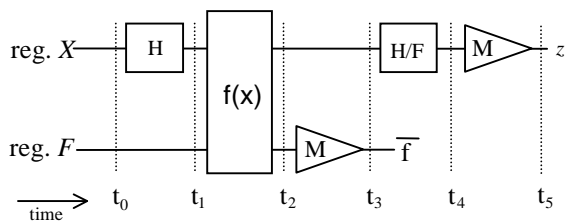
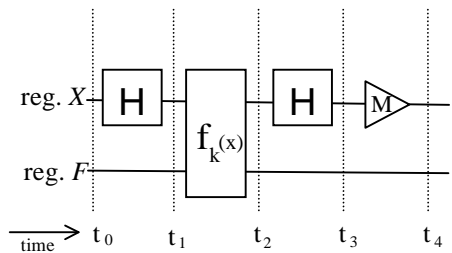
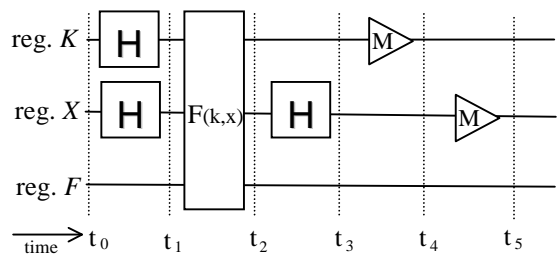
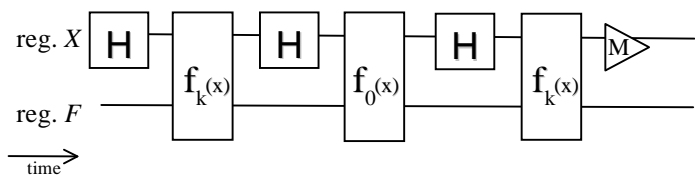
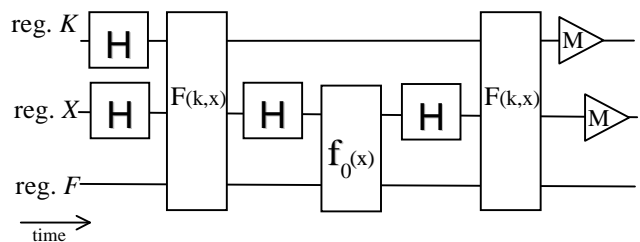
\end{document}